\documentclass{aastex62}

\def\teff{$T_\mathrm{eff}$}
\def\Msun{$\textit{M}_{\odot}$}
\def\Macc{$\dot{M}$}		
\def\Rsun{$\textit{R}_{\odot}$}
\def\Lsun{$\textit{L}_{\odot}$}
\def\Lbol{$\textit{L}_{bol}$}
\def\Lacc{$\textit{L}_{acc}$}
\def\Tbol{$\textit{T}_{bol}$}

\def\Vmax{$\textit{v}_{max}$}

\def\Tdisk{$\textit{T}_{disk}$}
\newcommand{\FeI}{\ion{Fe}{1}}
\newcommand{\FeII}{\ion{Fe}{2}}
\newcommand{\SII}{\ion{S}{2}}

\newcommand{\CaI}{\ion{Ca}{1}}

\newcommand{\NaI}{\ion{Na}{1}}
\newcommand{\kms}{km~s$^{-1}$}
\newcommand{\halp}{H$\alpha$} 
\newcommand{\hbet}{H$\beta$} 
\newcommand{\lam}{$\lambda$}

\newcommand{\um}{$\mu$m}
\newcommand{\hmol}{H$_{2}$}
\newcommand{\yr}{yr$^{-1}$}

\shorttitle{Spectra of V960~Mon}
\shortauthors{Park et al.}

\begin{document}

\title{High-resolution spectroscopic monitoring observations of FU Orionis-type object, V960~Mon}

\author{Sunkyung Park}
\affiliation{School of Space Research, Kyung Hee University \\
1732, Deogyeong-daero, Giheung-gu, Yongin-si, Gyeonggi-do, 17104, Republic of Korea \\ 
jeongeun.lee@khu.ac.kr, sunkyung@khu.ac.kr}

\author{Jeong-Eun Lee}
\affiliation{School of Space Research, Kyung Hee University \\
1732, Deogyeong-daero, Giheung-gu, Yongin-si, Gyeonggi-do, 17104, Republic of Korea \\ 
jeongeun.lee@khu.ac.kr, sunkyung@khu.ac.kr}

\author{Tae-Soo Pyo}
\affiliation{Subaru Telescope, National Astronomical Observatory of Japan \\
650 North A’ohoku Place, Hilo, HI 96720, USA}

\author{Daniel T. Jaffe}
\affiliation{Department of Astronomy, University of Texas at Austin \\
2515 Speedway, Austin, TX, USA}

\author{Gregory N. Mace}
\affiliation{Department of Astronomy, University of Texas at Austin \\
2515 Speedway, Austin, TX, USA}

\author{Hyun-Il Sung}
\affiliation{Korea Astronomy and Space Science Institute \\
776, Daedeok-daero, Yuseong-gu, Daejeon, 34055, Republic of Korea}

\author{Sang-Gak Lee}
\affiliation{Seoul National University \\
1 Gwanak-ro, Gwanak-gu, Seoul 08826, Republic of Korea}

\author{Wonseok Kang}
\affiliation{National Youth Space Center \\
200, Deokheungyangjjok-gil, Dongil-myeon, Goheung-gun, Jeollanam-do, 59567, Republic of Korea}

\author{Hyung-Il Oh}
\affiliation{Department of Astronomy and Atmospheric Sciences, Kyungpook National University \\
Daegu 41566, Republic of Korea}

\author{Tae Seog Yoon}
\affiliation{Department of Astronomy and Atmospheric Sciences, Kyungpook National University \\
Daegu 41566, Republic of Korea}

\author{Sung-Yong Yoon}
\affiliation{School of Space Research, Kyung Hee University \\
1732, Deogyeong-daero, Giheung-gu, Yongin-si, Gyeonggi-do, 17104, Republic of Korea \\ 
jeongeun.lee@khu.ac.kr, sunkyung@khu.ac.kr}

\author{Joel D. Green}
\affiliation{Space Telescope Science Institute \\
Baltimore, MD 21218, USA}

\begin{abstract}
We present the results of high-resolution (R $\ge$ 30,000) optical and near-infrared spectroscopic monitoring observations of a FU Orionis-type object, V960~Mon, which underwent an outburst in 2014 November.
We have monitored this object with the Bohyunsan Optical Echelle Spectrograph (BOES) and the Immersion GRating INfrared Spectrograph (IGRINS) since 2014 December. 
Various features produced by a wind, disk, and outflow/jet were detected.
The wind features varied over time and continually weakened after the outburst.
We detected double-peaked line profiles in the optical and near-infrared, and the line widths tend to decrease with increasing wavelength, indicative of Keplerian disk rotation. 
The disk features in the optical and near-infrared spectra fit well with G-type and K-type stellar spectra convolved with a kernel to account for the maximum projected disk rotation velocity of about 40.3$\pm$3.8~\kms\ and 36.3$\pm$3.9~\kms, respectively.
We also report the detection of [\SII] and \hmol~emission lines, which are jet/outflow tracers and rarely found in FUors.
\end{abstract}

\keywords{Stars: formation --- Stars: protostars  --- Stars: individual: V960~Mon, 2MASS~J06593168-0405277 --- Techniques: spectroscopic}

\section{Introduction} \label{sec:intro}
Low-mass stars form by gravitational collapse in dense molecular clouds.
The material is transferred from an infalling envelope to a disk and disk material is channeled into the central protostar along the magnetic field \citep{hartmann98}, growing the mass of the protostar.
However, the accretion mechanism from the disk to the central star is still poorly understood.
A steady accretion rate of $\sim$2 $\times$ 10$^{-6}$ \Msun~\yr~has been adopted in the standard accretion model \citep{shu77}. 
However, the observed luminosities of young stellar objects (YSOs) are often lower compared to the standard accretion model with the constant accretion rate, which is known as luminosity problem \citep{kenyon90, dunham10}.
A promising explanation of the luminosity problem is an episodic accretion model, where protostars spend most of their time in low accretion rates, and thus, with low luminosities, and occasional, relatively brief bursts of accretion dominate the time-averaged flow of material onto the central star and produce temporarily high luminosities and observable phenomena \citep{dunham10, audard14}.

FU Orionis-type objects (hereafter, FUors) are an observable evidence of episodic accretion; they are low-mass YSOs showing large-amplitude outbursts in the optical ($\Delta$V $\ge$ 4 mag) caused by a significantly enhanced mass accretion rate \citep[from 10$^{-7}$ to a few 10$^{-4}$ \Msun~\yr;][]{hartmann96, herbig03, hartmann09, audard14}.
In this context, FUors are ideal testbeds to study episodic accretion. 
During the outburst, the disk is about 100-1000 times brighter than the protostar, so the continuum source of the spectrum observed in optical and near-infrared (NIR) is the disk midplane \citep{hartmann96}. 
The major heating source of disk midplane is viscous heating caused by an accretion process \citep{armitage11}.

Originally, FUors were identified by their large brightness increase ($\Delta$V $\ge$ 4 mag) in the optical domain within a short timescale, and their label derives from their archetype FU Orionis \citep{herbig66}.
There have been many studies of FUors that sought to understand the accretion process in these outbursting objects.
\citet{herbig66, herbig77} organized the common phenomena and features of three FUors (FU Ori, V1057 Cyg, and V1515 Cyg), and those features have been used as criteria by which to identify FUors.
\citet{hartmann96} reviewed the properties of FUors and showed that an accretion disk model could explain their distinct features.
\citet{audard14} summarized the observational properties and theoretical interpretations of outbursts. 
Recently, \citet{connelley18} conducted a NIR spectroscopic survey of 33 FUors and FUor-like objects and presented the common NIR spectroscopic features of FUors.  
The characteristics of FUors are \citep{hartmann96, herbig03, hartmann09, audard14, connelley18}: (1) an increase in brightness by $>$ 4 mag in V band, (2) a very short rise time (1-10~yr) followed by a long decay time (from decades to centuries), (3) association with distinctive reflection nebulae, (4) strong infrared excesses, (5) \textquotedblleft double-peaked\textquotedblright~metallic absorption profiles, (6) broad blue-shifted Balmer lines, (7) P Cygni profiles of \halp~and \NaI~D~lines, (8) strong CO absorption features, (9) strong water absorption bands at the edges of the H band, (10) a strong blue-shifted He I absorption line profile, and (11) wavelength-dependent spectral types; optical and infrared spectra are consistent with F-G and K-M supergiant or giant, respectively.

There are only about 30 known FUors and FUor-like objects.
Low-resolution optical and NIR spectra are a powerful tool with which to identify FUors \citep{hartmann96}.
With the addition of high-resolution data, we can study the physical and kinematic structure of the inner disk. 
To date, HBC 722 is the only FUor that has been studied with high-resolution (R~$\equiv$~\lam/$\Delta$\lam~$\ge$~30,000) optical and NIR spectroscopic monitoring \citep{lee15}.
\citet{lee15} detected strongly blue-shifted absorption profiles formed by a wind and broad double-peaked absorption profiles originating in the disk of HBC~722 and found that the relative strengths of these two types of lines are anti-correlated in time.
\citet{lee15} conclude that this anti-correlation arises because wind pressure in the early phase of the outburst would prevent the rebuilding of the inner hot gaseous disk.
Moreover, with the high-resolution spectra, \citet{lee15} were able to distinguish the protostellar lines from lines produced by the disk with its heated midplane (see their Fig.~2). 
The HBC~722 results highlight the importance of high-resolution spectroscopic studies in understanding the physical processes associated with an accretion burst.

V960 Mon (2MASS~J06593158-0405277) exhibited an outburst in 2014 November and has been identified as a FUor \citep{maehara14, hillenbrand14}.
V960~Mon is located towards the Lynds 1650 cloud \citep{reipurth15, pyo15}.
The calculated distance using the Gaia DR2 \citep{gaia18} data is about 1638$\pm$163~pc.
The optical and NIR spectra show characteristics of FUors \citep{hillenbrand14, reipurth15, pyo15, kospal15, caratti15, takagi18}.
While there are intentionally taken pre-outburst data for only two FUors: V1057 Cyg \citep{herbig77} and HBC 722 \citep{cohen79}, V960~Mon lies close to the Galactic plane so Galactic plane surveys can furnish information about the pre-outburst stage \citep{kospal15}.
Therefore, the V960~Mon dataset provides the opportunity to investigate the entire FUor phenomenon from its pre-outburst to its post-outburst phase.

\citet{kospal15} studied the properties of the pre-outburst stage of V960~Mon. 
They estimated the central protostellar temperature as about 4000~K and central protostellar mass as about 0.75 \Msun.
From the estimated properties, V960~Mon was classified as Class II before the outburst.
\citet{hackstein15} suggested an oscillating period of 17 days from the post-outburst light curve.
\citet{caratti15} revealed the existence of an extended disk-like structure, a very low-mass companion, and a bump which is thought to be a closer companion.
They suggest that the target is a triple system, and the observed NIR spectral features of V960~Mon are similar to those of HBC~722 \citep{miller11}.
\citet{jurdana16} re-constructed the historical light curve from 1899 to 1989 but did not find any brightness change similar to the outburst that occurred in 2014.
Recently, \citet{takagi18} presented observations of spectroscopic variations of \halp~and nearby atomic lines and suggested that these variations are caused by the decreasing mass accretion rate.

In this paper, we present high-resolution optical and NIR spectroscopic monitoring observational results for V960~Mon and suggest that this FUor was in the earliest stage of the Class II phase prior to its outburst.

\section{Observations and Data Reduction} \label{sec:observation}
\subsection{Optical Observations}
We observed V960~Mon using BOES from 2015 February 11 to 2018 December 19 with a spectral resolution of 30,000 using a 300~\um~fiber.
BOES is an echelle spectrograph \citep{kim02} attached to the 1.8~m optical telescope at Bohyunsan Optical Astronomy Observatory (BOAO) in Korea, and it covers the full optical wavelength from 3,900~\AA{} to 9,900~\AA{}. 
To increase the signal-to-noise ratio (S/N), we binned the spectra over 2 $\times$ 2 pixels.
The S/N around 6000~\AA{} is typically 30, and ranges from 17 to 44. 
Standard stars were also observed with the same observational setup as V960~Mon to aid in the spectral analysis.
Fig.~\ref{fig_lightcurve} shows the light curve of V960~Mon and black dotted lines indicate the observation dates of BOES.
Table~\ref{tbl_obs_tot} lists the observing log.

The observed spectra were reduced using the IRAF \citep{tody86, tody93} {\tt echelle} package. 
For each image, bias subtraction was conducted, and each aperture from the spectral images was extracted using a master flat-field image.
As part of the flat-fielding process, we corrected the interference fringes and pixel-to-pixel variations of the spectrum images.
A ThAr lamp spectrum was used for wavelength calibration. 
Continuum fitting was performed by the {\tt continuum} task.
Finally, heliocentric velocity correction was applied by using the {\tt rvcorrect} task and the published radial velocity of V960~Mon \citep[38.1$\pm$0.5~\kms;][]{takagi18}.

\subsection{Near-infrared Observations} 
We observed V960~Mon with IGRINS installed on the 2.7~m Harlan J. Smith Telescope (HJST) at McDonald Observatory and on the 4.3~m Discovery Channel Telescope (DCT) at Lowell Observatory from 2014 December 25 to 2017 November 26.
IGRINS provides high-resolution (R~$\sim$~45,000) NIR spectra covering the full H~(1.49-1.80~\um) and K~(1.96-2.46~\um) bands with a single exposure \citep{yuk10, park14}.
Table~\ref{tbl_obs_tot} lists the observing log for IGRINS and gray dashed lines in Fig.~\ref{fig_lightcurve} indicate the dates of IGRINS observations.

We reduced the H and K spectra using the IGRINS pipeline \citep{lee17} for flat-fielding, sky subtraction, correcting the distortion of the dispersion direction, wavelength calibration, and combining the spectra.
Telluric standard stars (A0~V) were observed immediately after or before each observation of V960~Mon for telluric correction.
Continuum fitting and telluric correction were performed using custom IDL routines.
We applied the same method for the entire data reduction as described in \citet{park18}. 
We report a S/N for each spectrum based on the median value in the order that covers from 2.21 to 2.24~\um.
The S/N is typically 190, and ranges from 93 to 289.
Finally, a heliocentric velocity correction was applied using the same method as used for the optical spectra.

\section{Results and Analysis} \label{sec:results}

\subsection{Wind Features} \label{subsec:wind}
The mass loss rate for Class II YSOs is about 10$\%$ of the mass accretion rate \citep{hartmann96, hartmann09, ellerbroek13, bally16}.
FUors have more powerful winds than other YSOs because their mass accretion rates ($\sim$~10$^{-5}$ to 10$^{-4}$ \Msun~\yr) are about three orders of magnitudes greater than other Class II YSOs \citep[$\sim$~10$^{-8}$ to 10$^{-7}$ \Msun~\yr;][]{hartmann96, herbig03, hartmann09, audard14, hartmann16}. 
During the outburst, high-velocity winds (several hundred \kms) can be present \citep{hartmann96}.
\citet{calvet93} and \citet{hartmann96} showed that wind features can arise from the accreting disk.
The stronger wind lines are formed at the vertically outer part of the disk atmosphere, which indicate the largest expansion velocities and show strongly blue-shifted absorption profiles.

The optical spectra of the V960~Mon show several wind features in lines of \hbet~4861~\AA{}, \NaI~D doublet (5889~\AA{} and 5895~\AA{}), and \halp~6563~\AA{} (Fig.~\ref{fig_wind}).
All of these transitions of wind features have blue-shifted absorption profiles \citep{bastian85, herbig03, hartmann09} that appear to vary with time. 
The \halp~and \hbet~line profiles, in particular, show clear changes.

\halp~has a P~Cygni~profile with a strong and broad blue-shifted absorption component extending to about -400~\kms, which is produced by an outflowing wind \citep{hartmann96, hartmann09, herbig09, reipurth10, lee11}. 
A red-shifted emission component is also present. 
The variation of the blue-shifted absorption component with time is significant: the depth was the most profound at the first observation (2015 February, shortly after the outburst) and became shallower until 2015 October.
About one year after the outburst, the absorption component disappeared and the blue-shifted side of the line was in emission after 2015 December (Fig.~\ref{fig_wind}).
At the same time, the width of the blue-shifted absorption features of \hbet~and \NaI~D~doublets in Fig.~\ref{fig_wind} became narrower and shallower with time. 
Fig.~\ref{fig_Halp_Hbet} shows line variations of \hbet~(left panel) and \halp~(right panel) with time. 
As shown in Figs.~\ref{fig_wind} and~\ref{fig_Halp_Hbet}, the depth variation of the broad absorption component of \halp~and the width variation of \hbet~occur simultaneously.
These changes in wind features imply that the blue-shifted component is continually weakening since our observation, and the high-velocity component of the wind became too weak to be detectable around 2015 December. 
The changes in the blue-shifted component of wind features can be explained by the decreasing mass accretion rate.

\subsection{Disk Features} \label{subsec:disk}
A disk in Keplerian rotation can produce double-peaked line profiles whose peak separations decrease with increasing wavelength \citep{hartmann96, hartmann09, zhu07, zhu09}. 
In addition to the double-peaked line profile, a Keplerian rotational disk can produce a boxy profile with a flat bottom and steep wings \citep{petrov08}.

\begin{figure*}
\plotone{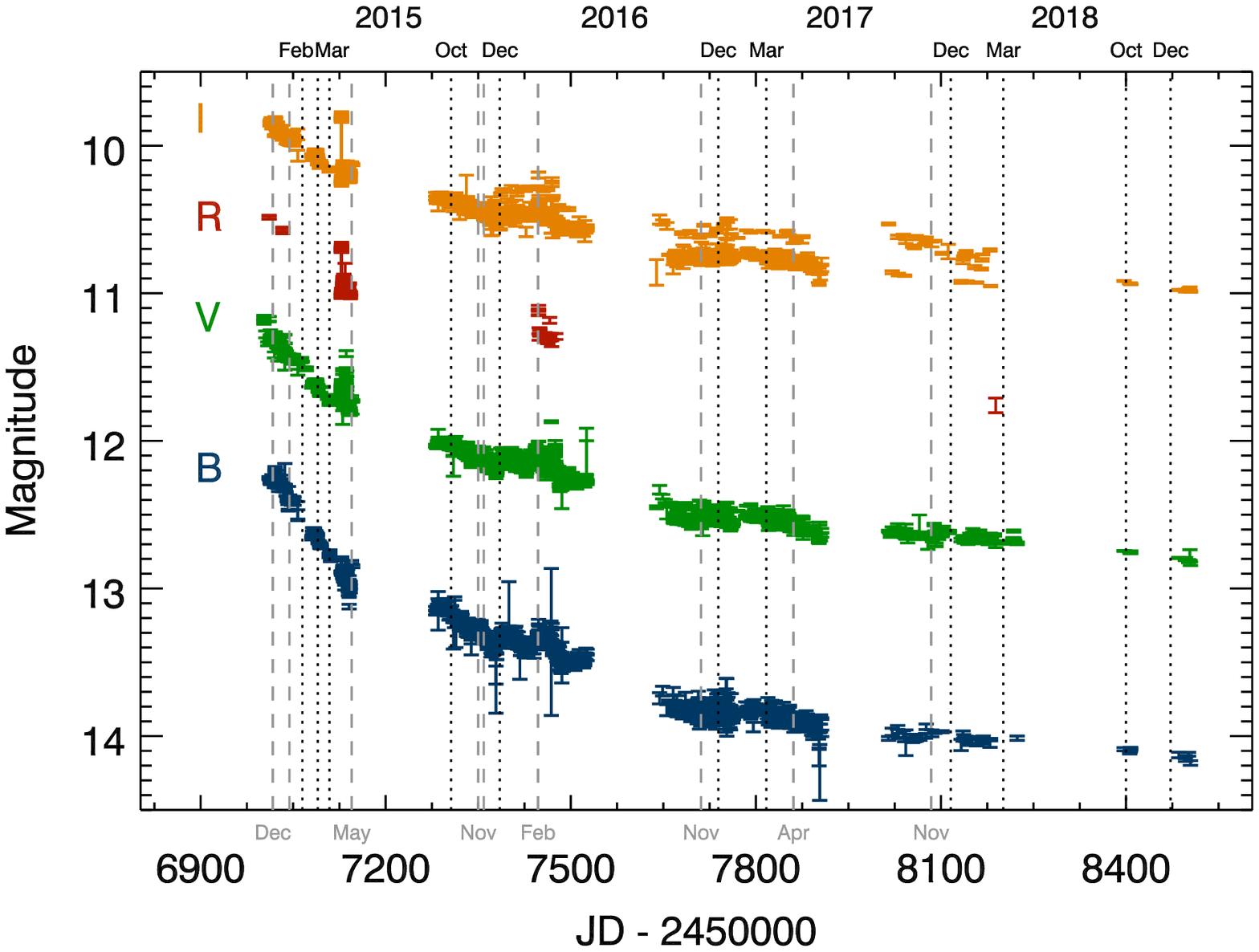} 
\caption{Light curve of V960~Mon.
The brightness of V960~Mon decreases after its outburst occurred in 2014 November.
Gray dashed and black dotted lines indicate the observation dates of IGRINS and BOES, respectively.
The numbers and characters in the top x-axis indicate the calendar date of the observed year and month for BOES.
The small gray characters in the x-axis indicate the calendar date of the observed month for IGRINS. 
The BVRI data are from the AAVSO data archive (\url{https://www.aavso.org}).
Different colors represent different bandpasses.
\label{fig_lightcurve}
}
\end{figure*}

\begin{figure*}
\centering
\epsscale{1.25}
 \includegraphics[width=0.9\textwidth]{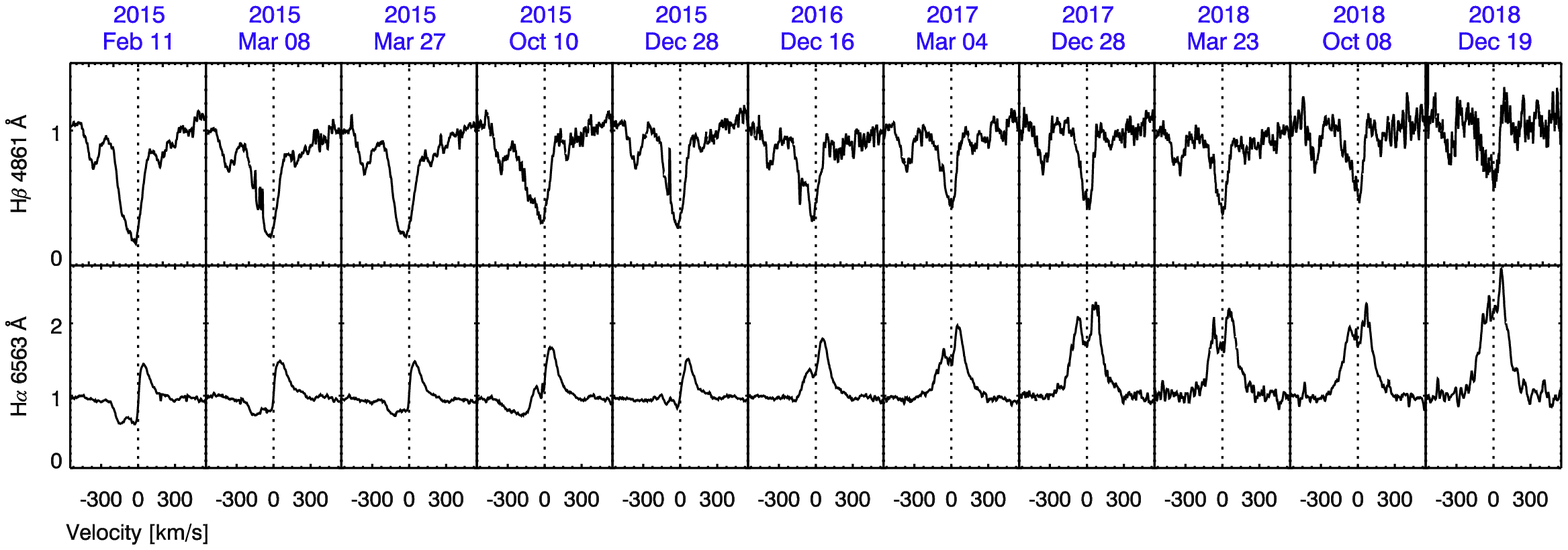} 
 \includegraphics[width=0.9\textwidth]{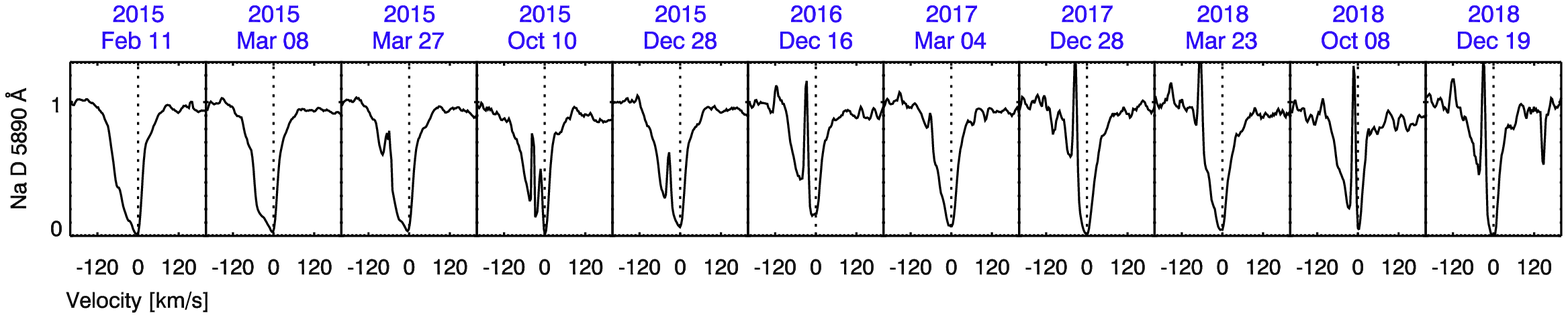}
\caption{
Time variation of wind features (top: \hbet~4861~\AA{} and \halp~6563~\AA{}, bottom: Na~D~5890~\AA{}).
Different wind features are shown in rows, and different observation dates are shown in columns.
The depth and strength of blue-shifted absorption profiles of all wind features are the strongest at the first observation (2015 February 11) and became narrower and shallower with time.
Narrow emission lines superposed on the absorption lines of Na~D~5890~\AA{} are sky emission lines \citep{hanuschik03}.
\label{fig_wind}
}
\end{figure*}

\begin{figure*}
\centering
\plottwo{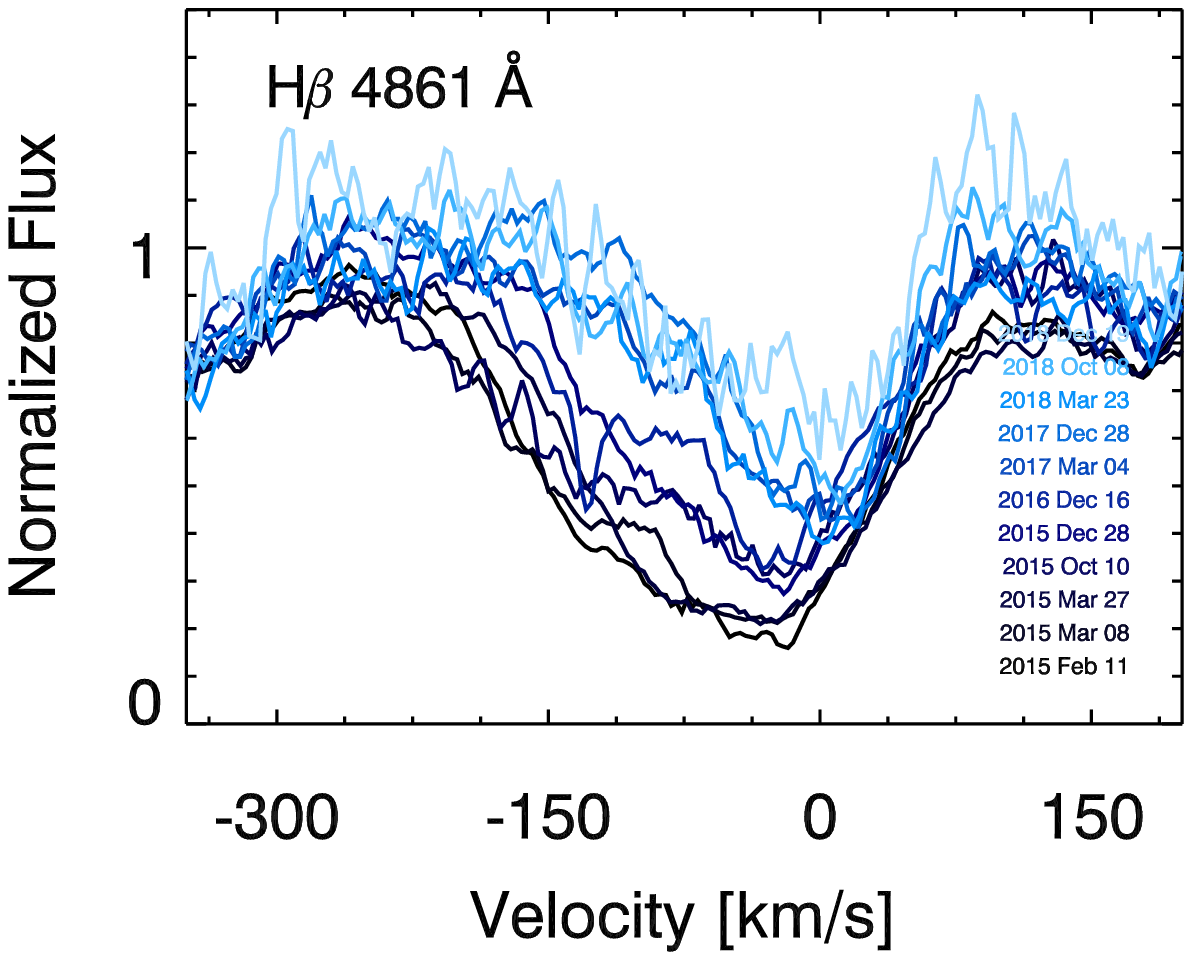}{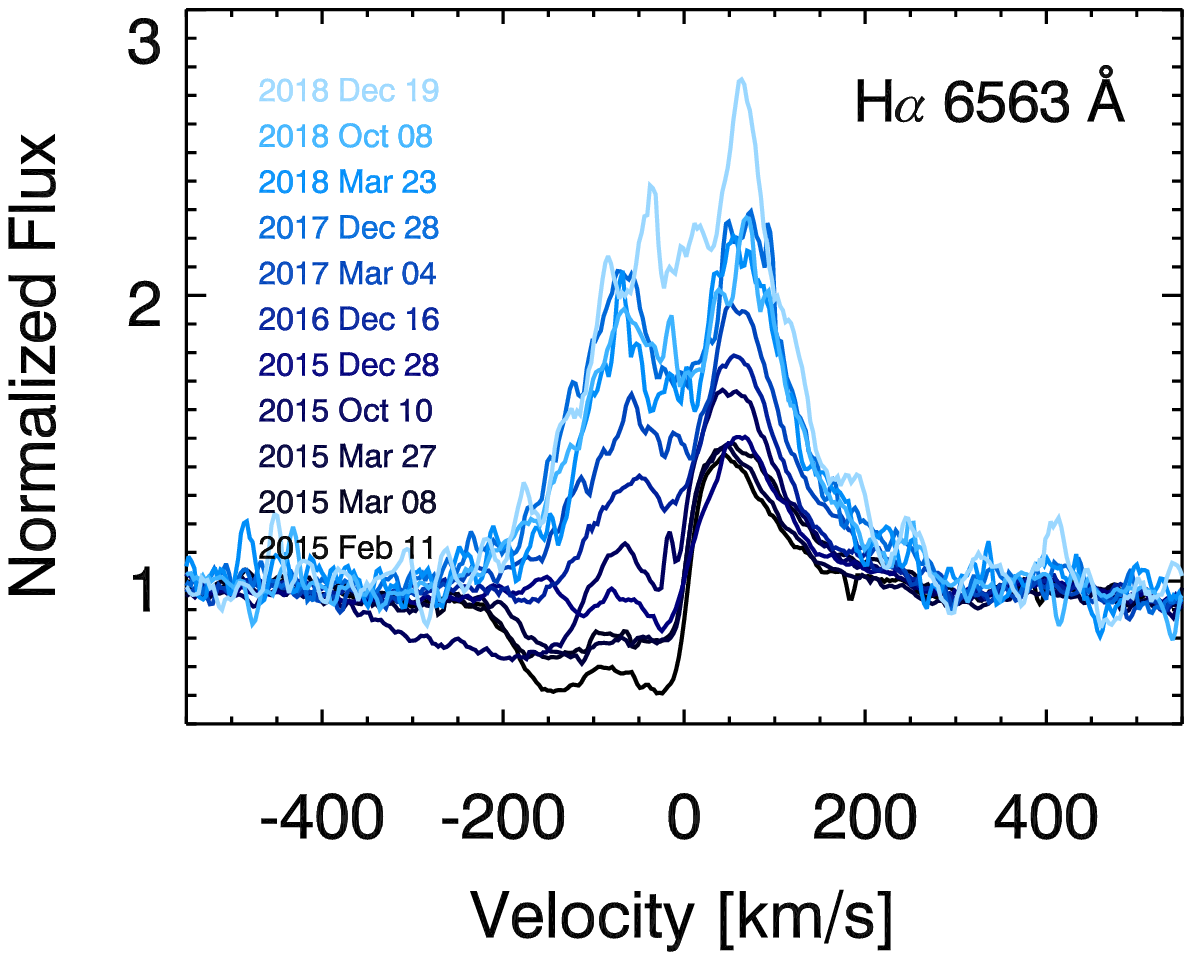}
\caption{Time variation of \hbet~4861~\AA{} (left) and \halp~6563~\AA{} (right).
Different colors indicate different observational dates.
About one year after its outburst (between 2015 October and 2015 December), blue-shifted absorption component of P~Cygni profile disappeared in \halp. 
Then, the blue-shifted absorption component of \halp~turned into emission.
In the same time, blue-shifted absorption features of \hbet~became narrower and shallower. 
\label{fig_Halp_Hbet}
}
\end{figure*}

\begin{figure}
\epsscale{0.5}
\centering
\plotone{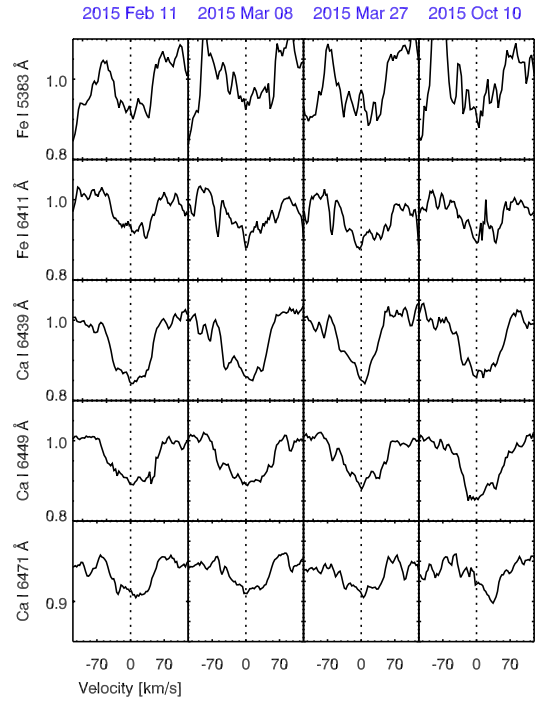}
\caption{
Boxy or double-peaked absorption line profiles of optical spectra.
Different disk features are shown in rows, and different observation dates are shown in columns.
The SNR was the highest at the first observation (2015 February 11) right after the burst, and the SNR decreases due to continuous brightness decrease.
Therefore, the first four observation data were used for analysis.
\label{fig_disk_boes}
}
\end{figure}

\begin{figure}
\epsscale{1.1}
\centering
\plotone{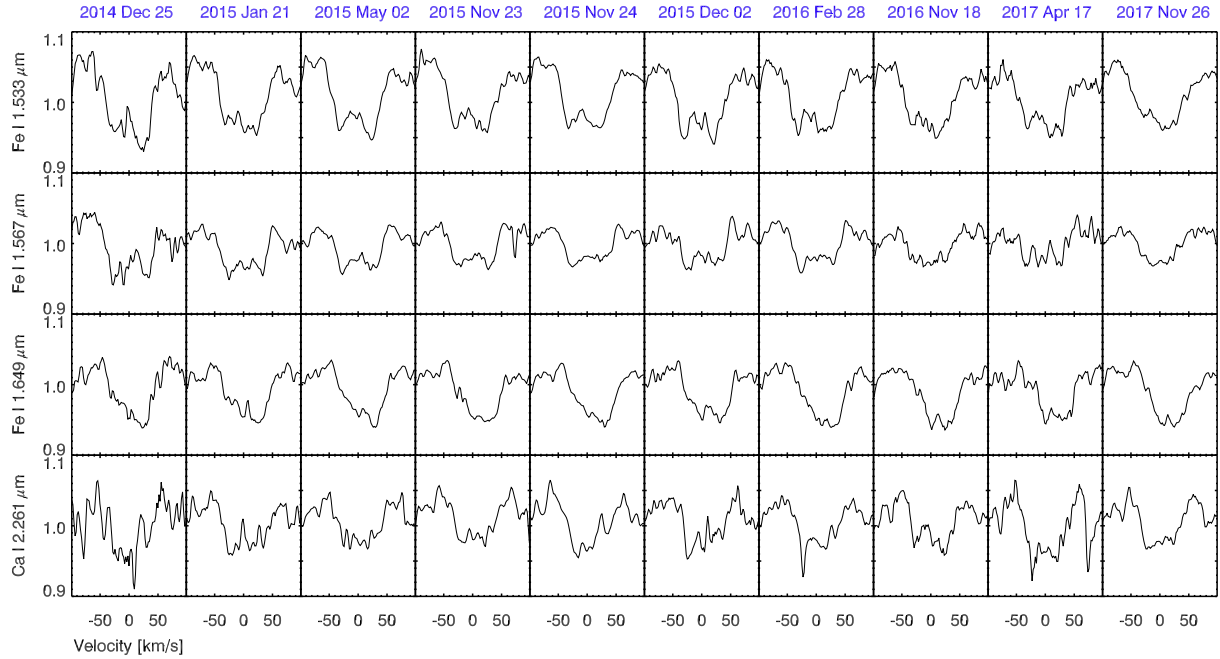}
\caption{
Boxy or double-peaked absorption line profiles of NIR spectra.
Different disk features are shown in rows, and different observation dates are shown in columns.
\label{fig_disk_igrins}
}
\end{figure}

Several atomic metal lines were detected with double-peaked or boxy profiles in optical and NIR spectra (Fig.~\ref{fig_disk_boes} and Fig.~\ref{fig_disk_igrins}).
The double-peaked lines are relatively clear at the first observations (2015 February for BOES and 2014 December for IGRINS), consistent with wind features (Section 3.1).
The S/N is the best at the first observation dates for each observation of BOES and IGRINS because the source was at its brightest immediately after the outburst and became fainter with time \citep[Fig.~\ref{fig_lightcurve} and][]{hackstein15}.
The optical and NIR spectra of FUors during outburst originate from the disk rather than the central star because of the significant mass accretion rate \citep{petrov92, hartmann96}.
The dimming results from the decreasing mass accretion rate after the outburst (Fig.~\ref{fig_lightcurve}), which reduces continuum brightness of the disk midplane.
Therefore, only the first few observational data  are used for disk analyses in this section.

If V960~Mon has a Keplerian disk with a radially decreasing temperature, the longer wavelength traces the larger radius where the disk rotates more slowly. 
We fit these double-peaked line profiles by convolving standard stellar spectra with a disk rotational profile and estimated the temperature and radius where the observed lines are formed.
The standard stellar spectra are convolved with a disk rotational profile as below \citep{calvet93, hartmann96, hartmann09}.
\begin{equation} \label{eq_rconvol}
\phi(\Delta{v})=\Big[1- \big(\frac{\Delta{v}} {v_{max}}\big)^{2}\Big]^{-1/2}, 
\end{equation}
where $\Delta{v}$ is the velocity shift from the line center and $v_{max}$ is the maximum projected rotational velocity ($v_{max} = vsini$).

\begin{figure}
\epsscale{1.1}
\plottwo{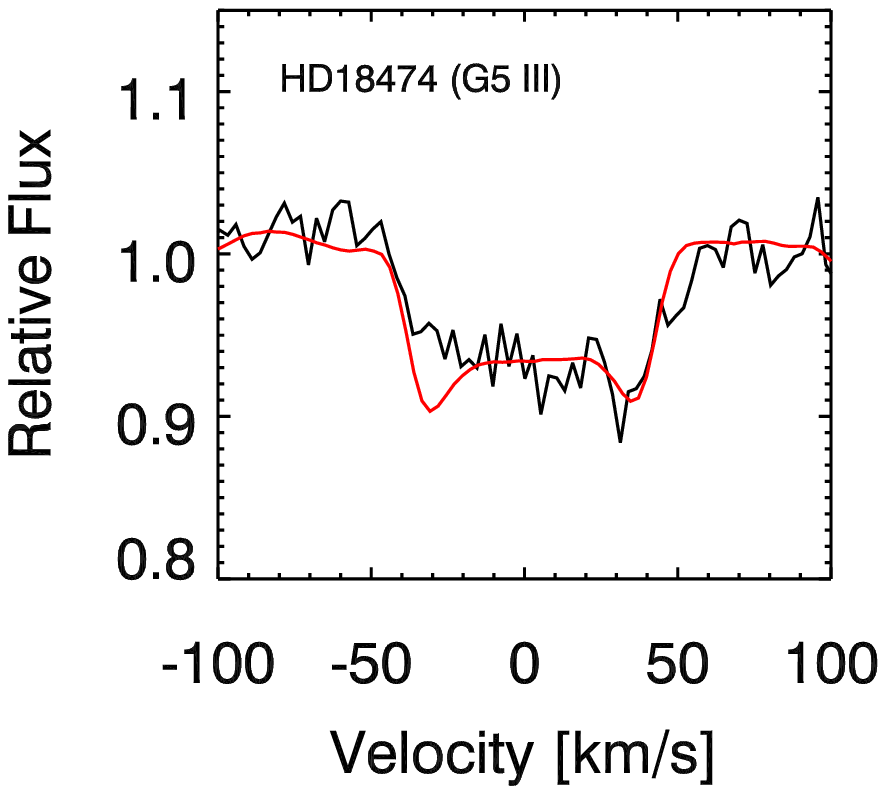}{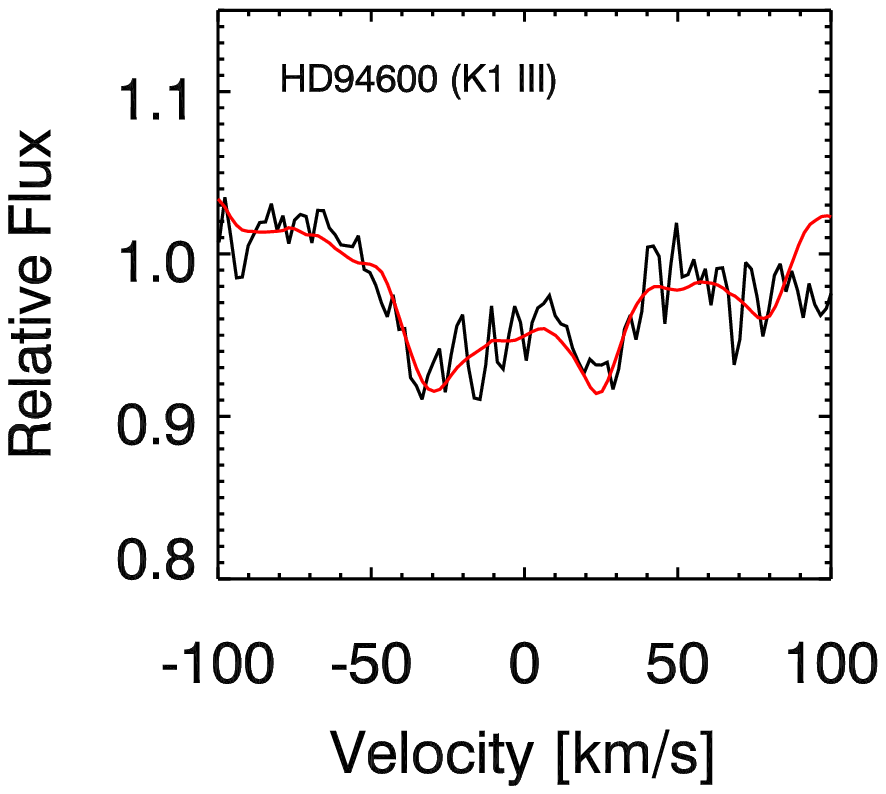}
\caption{Double-peaked line profiles of \FeI~6411~\AA{} (left) and \FeI~1.566~\um~(right).
Black lines present the disk features of V960~Mon.
Red lines indicate the stellar spectra convolved with disk rotational profile.
Optical spectrum (left) fit well with the G5-type (HD~18474) stellar spectrum convolved with a projected rotational velocity of 44$\pm$1~\kms, while NIR spectrum (right) fit well with the K1-type (HD~94600) stellar spectrum convolved with a projected rotational velocity of 32$\pm$0.5~\kms. 
The rotational velocity of the double-peaked lines decreases with increasing wavelengths; optical spectrum trace warmer inner part of the disk and NIR spectrum trace cooler outer part of the disk.
\label{fig_convol}
}
\end{figure}

For the optical analysis, we observed several standard stars with the same observational setup as V960 Mon.
We performed disk rotational convolution in steps of 1~\kms\ and obtained consistent results within 3~\kms\ intervals, therefore, the uncertainty of the fitting is $\pm$~1~\kms. 
The best-fit was determined by the chi-square minimization from the fitting of the double peak/boxy lines, and the spectra of HD~219477 (G2~II-III) and HD~18474 (G5~III) fit the best the spectra of V960~Mon.
The best-fit results of optical double-peaked/boxy lines are found in the ranges of 35-44~\kms, and the average and standard deviation of \Vmax\ is 40.3$\pm$3.8~\kms.
However, there is additional uncertainty in the fitting results because of the coarse grid of spectral types and luminosity classes for standard stars.
Fig.~\ref{fig_convol} shows two example spectra of the best-fit results for each of the optical and NIR. 
We adopted the~\teff~of HD~18474 (G5~III) as 5013~K from \citet{liu14}.
In the case of HD~219477 (G2~II-III), the \teff~was unknown. 
Therefore, we calculated the~\teff~as about 5300~K by adopting \teff-(B-V) relation \citep{flower96, torres10}.

In analyzing the NIR spectra of V960~Mon, we used the spectra of standard stars from the IGRINS Spectral Library \citep{park18}, and most of the double-peaked lines were fitted well by a K1-type \citep[HD~94600 (K1~III), \teff~$\sim$~4600~K;][]{wu11} stellar spectrum. 
The disk rotational convolution was performed in steps of 1~\kms\ and obtained consistent results within 2~\kms\ intervals, therefore, the uncertainty of the fitting is $\pm$~0.5~\kms.
The best-fit results of NIR double-peaked/boxy lines are found in the ranges of 32.5-41.5~\kms, and the average and standard deviation of \Vmax\ is 36.3$\pm$3.9~\kms.

If V960~Mon has a Keplerian disk with an inclination of 90, and the central protostellar mass of 0.75~\Msun~\citep{kospal15}, the observed double-peaked optical and NIR lines trace 88$\pm$32~\Rsun\ and 109$\pm$40~\Rsun\ of the disk, respectively.
Hence, the temperature of the disk decreases from 5300-5000~K at 88$\pm$32~\Rsun~to 4600~K at 109$\pm$40~\Rsun.
These results show that the disk features at the longer wavelengths trace the cooler outer part of the disk with lower rotational velocity.
The estimated disk radii depend on the inclination of V960 Mon, and the detected disk features show double-peaked absorption profiles, which indicate that the disk inclination is less than 90 degrees. 
According to \citet{caratti15}, the inclination of the disk-like structure is about 22 (28) degrees, which results in the optical and NIR spectra tracing 12$\pm$4~(19$\pm$7)~\Rsun\ and 15$\pm$6~(24$\pm$8)~\Rsun, respectively.

\begin{figure}
\epsscale{0.7}	
\plotone{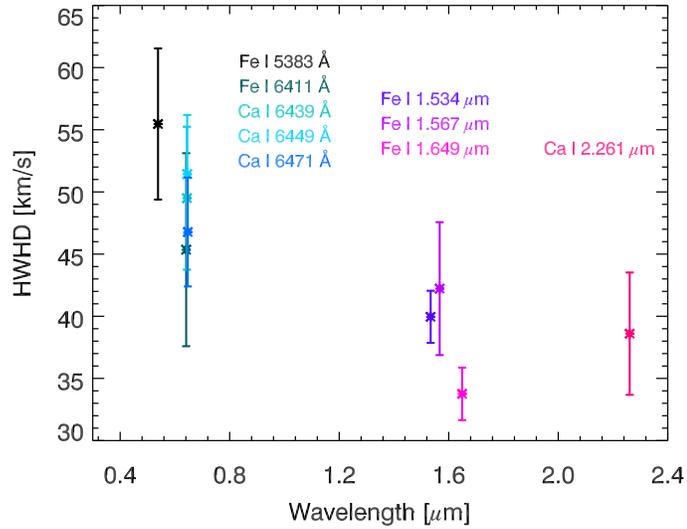} 	
\caption{HWHD of double-peaked absorption features as a function of wavelength. 
Different colors present different lines.
The HWHD becomes narrower as the wavelength becomes longer, which is consistent with the Keplerian disk rotation. 
\label{fig_hwhd}
}
\end{figure}

We measured the Half-Width at Half-Depth \citep[HWHD,][]{petrov08} of the \FeI~5383~\AA{}, \FeI~6411~\AA{}, \CaI~6439~\AA{}, \CaI~6449~\AA{}, \CaI~6471~\AA{}, \FeI~1.534~\um, \FeI~1.567~\um, \FeI~1.649~\um, and \CaI~2.261~\um~lines and list the measured values in Table~\ref{tbl_hwhd}. 
As discussed above, we measured the HWHD of each line only the first few observations for BOES (from February 2015 to October 2015) and all observations for IGRINS and plot the averages and standard deviations in Fig.~\ref{fig_hwhd}.
The average HWHD of optical and NIR lines are about 50$\pm$6~\kms~and 39$\pm$4~\kms, respectively.  
The HWHD decreases with increasing wavelength, consistent with the origin in a Keplerian disk where hotter inner material is rotating faster than cooler outer material.

\begin{figure}
\epsscale{1.}
\plotone{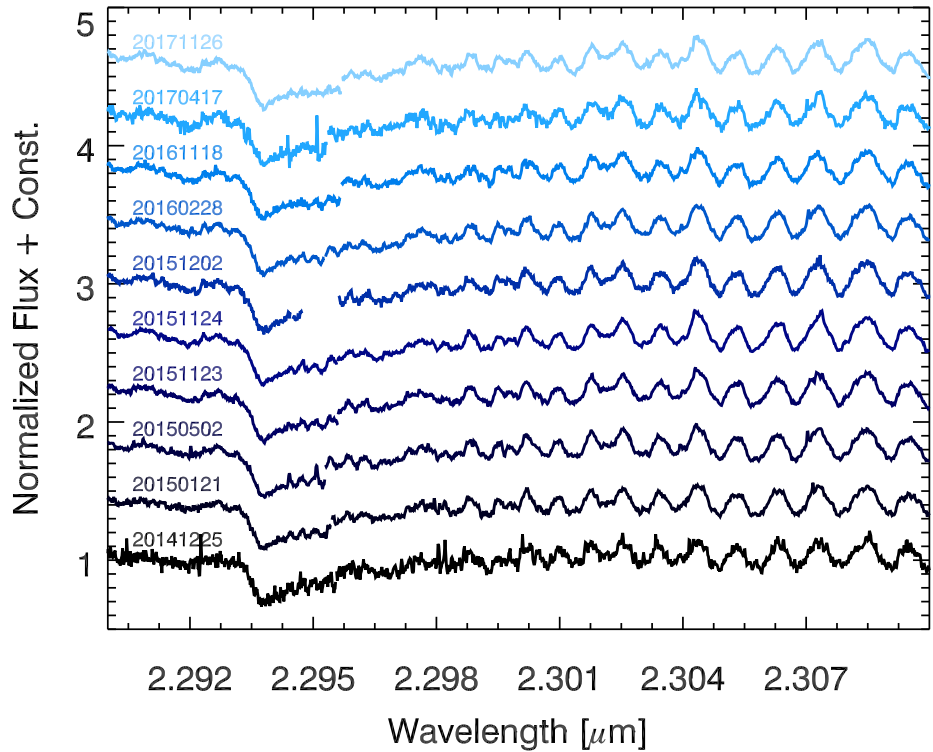}
\caption{
The CO first overtone band transitions of V960~Mon observed with IGRINS.
Different colors indicate different observation dates.
There is no significant change in CO absorption features.
\label{fig_co_variation}
}
\end{figure}

\begin{figure}
\epsscale{1.}
\plotone{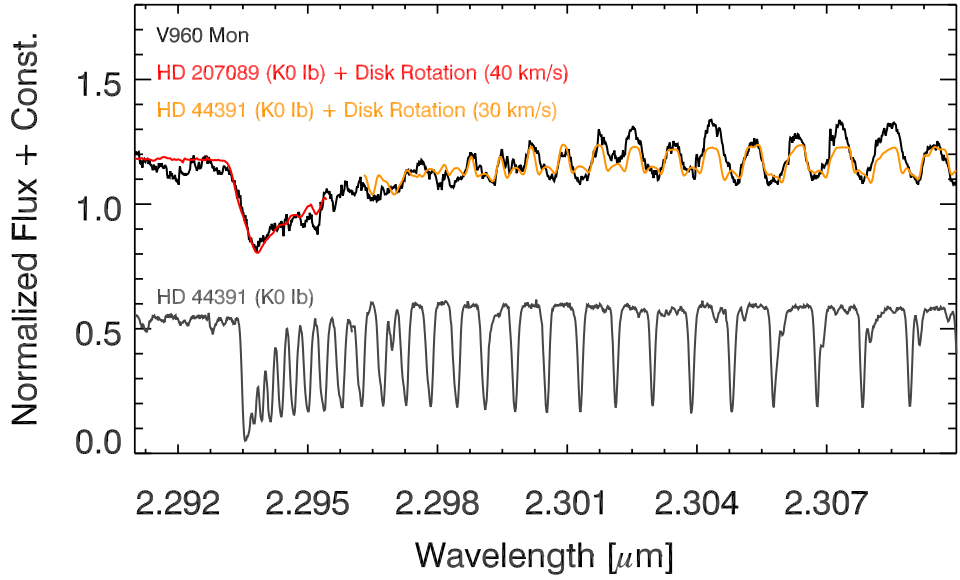} 
\caption{
The CO overtone transitions in V960~Mon (black) and standard star HD~44391 (K0~Ib; gray).
The best-fit stellar spectrum of HD~207089 (K0~Ib; red) and HD~44391 (orange) convolved with a disk rotational profile of 40~\kms\ and 30~\kms\ is presented, respectively.
The spectral features of V960~Mon are much broader than those of standard star, while they are reasonably matched with the stellar spectra convolved with a disk rotational profile.
\label{fig_co}
}
\end{figure}

In addition, the IGRINS spectra clearly show strong CO absorption features at 2.293 \um~(Fig.~\ref{fig_co_variation}), one of the representative characteristics of FUors \citep{hartmann96, audard14, connelley18}, which are produced against the heated midplane by the accretion burst, and the broadened CO features are caused by the Keplerian rotation.
There is no significant variation in CO absorption features during our NIR observations.
V960~Mon shows broader line widths (black line in Fig.~\ref{fig_co}) than the standard star \citep[gray line; HD~44391,][]{park18} because of the disk rotation. 
The CO absorption features are reasonably well fitted, but not perfectly, by the stellar spectrum of HD~207089 and HD~44391 convolved with a projected rotational velocity of 40~\kms\ and 30~\kms\ for bandhead (red) and rovibrational lines (orange), respectively.
This result suggests that the lower energy transitions of the CO overtone band are produced at larger radii where the disk rotates more slowly.

\subsection{Outflow/Jet Features} \label{subsec:wind}
\begin{figure}
\epsscale{0.8}
 \includegraphics[width=8.75cm, height=7cm]{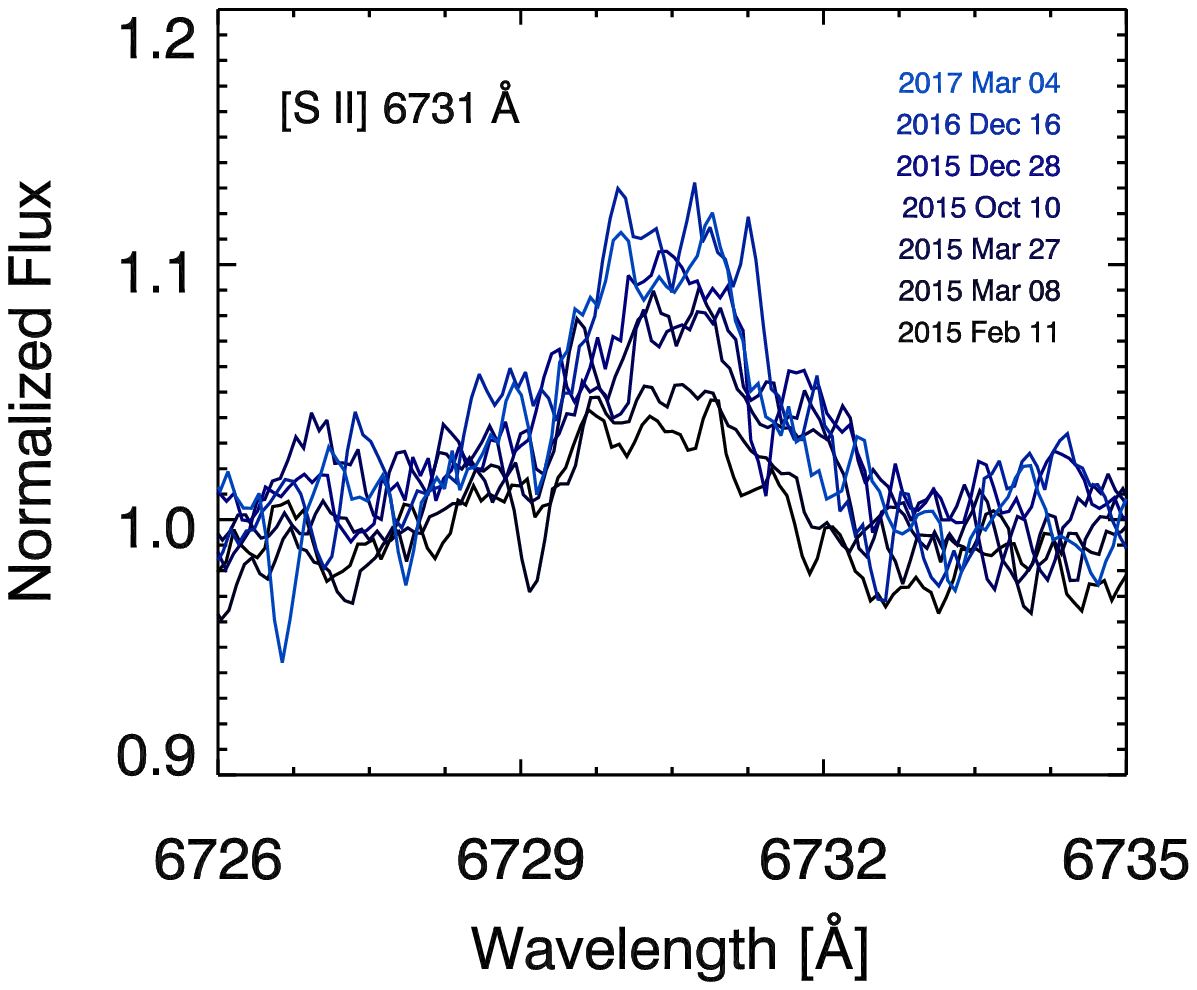}
 \includegraphics[width=8.75cm, height=7cm]{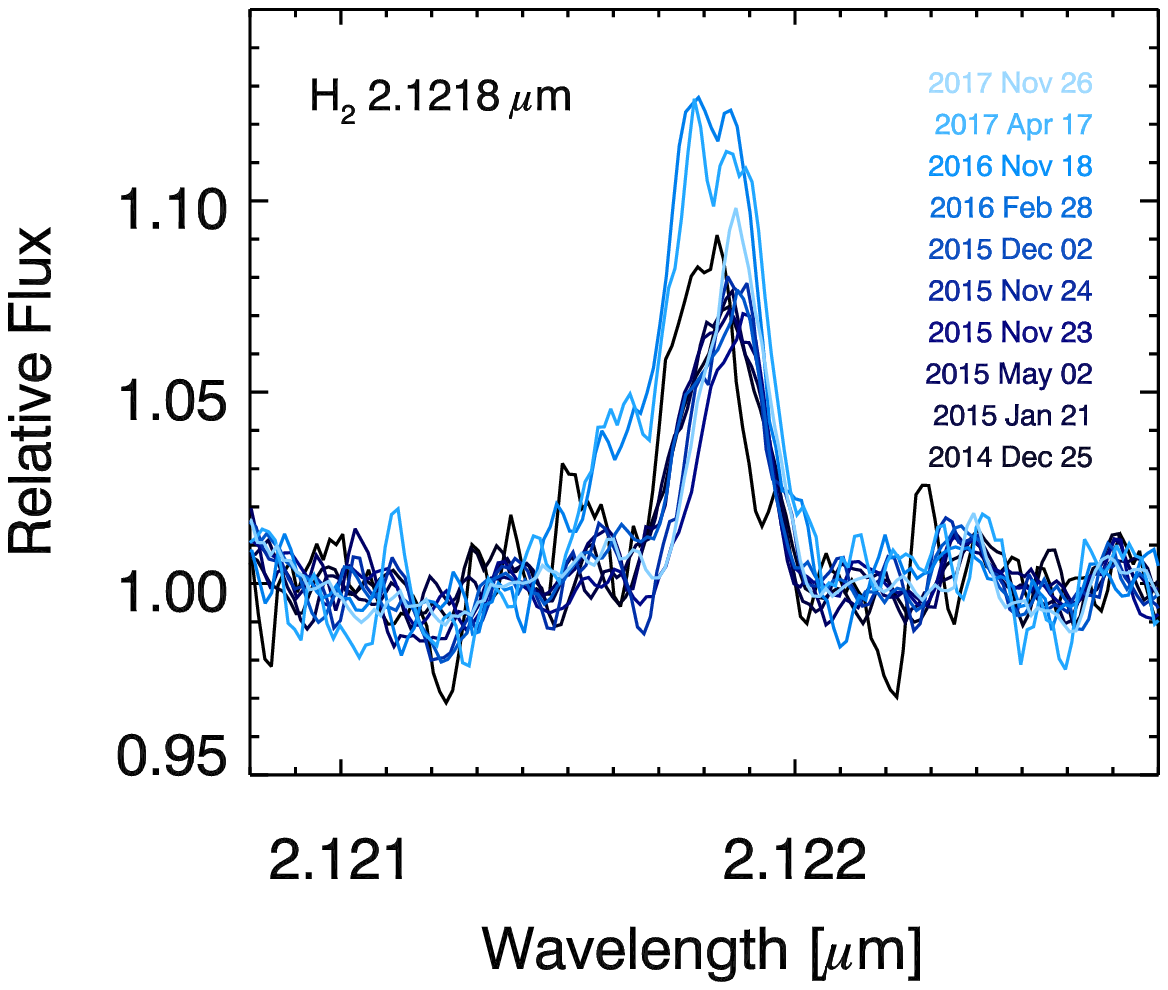}
\caption{Emission lines of V960~Mon.
Left and right panels show the [\SII]~6731~\AA{} and the \hmol~2.1218~\um~emission lines, respectively.
Different colors present different observation dates.
The [\SII]~6731~\AA{} emission line was shown until 2017 March, because of the low S/N ($\le$ 10) since 2017 December.
The [\SII]~6731~\AA{} line (FWHM $\sim$ 106$\pm$9 \kms) is broader than the \hmol~2.1218~\um~line (FWHM $\sim$ 27$\pm$6 \kms).
\label{fig_emission}
}
\end{figure}

Emission lines are, in general, hardly detected in FUors, except \halp~P~Cygni profile.
The optical and NIR spectra of V960~Mon show emission lines of [\SII]~6731~\AA{} and \hmol~2.1218~\um~(Fig.~\ref{fig_emission}). 
The mean S/N of the [\SII]~6731~\AA{} and the \hmol~2.1218~\um\ line are about 69 and 191, respectively.
\citet{takagi18} also detected the [\SII]~6731~\AA{} emission line in V960~Mon.
Before the detection of emission lines in V960~Mon, V2494~Cyg was the only FUor that showed emission line of the [\SII]~6731~\AA{} \citep{magakian13}. 
The \hmol~2.1218~\um\ spectrum in Fig.~\ref{fig_emission} represents the first detection of this feature in a FUor spectrum.

The [\SII]~6731~\AA{} emission line is a well-known outflow/jet tracer in Class II objects \citep{hirth97, simon16}.
According to \citet{hartmann09}, the [\SII]~6731~\AA{} emission line can be formed in the entrained gas accelerated by a highly collimated jet.
The peak velocity of the [\SII]~6731~\AA{} emission line is blue-shifted with respective to the systemic velocity by 19$\pm$4~\kms.
The [\SII]~6731~\AA{} emission line has similar physical properties to the [\FeII]~1.644~\um, but the [\FeII] line has a higher critical density ($\sim$3~$\times$ 10$^{4}$ cm$^{-1}$) than that of the [\SII] ($\sim$2 $\times$ 10$^{3}$ cm$^{-1}$) \citep{reipurth00, nisini05, hayashi09}.
Since only the [\SII]~6731~\AA{} emission was detected, we infer that the outflow has a lower density.

The \hmol~2.1218~\um~emission line is also known as a tracer of outflows at the earlier stages of YSOs, in particular, Class~I \citep{davis03, bally07, davis10, greene10, bally16}. 
Generally, the [\FeII]~1.644~\um\ emission line arises from fast shock ($>$ 30 \kms) while the \hmol~2.1218~\um~emission line arises from relatively slower shock \citep[$<$ 25 \kms;][]{hayashi09}; the mechanism of line formation mainly depends on the shock velocity.
In the spectra of V960~Mon, only the \hmol~emission line was detected, and its peak velocity is 5$\pm$4~\kms.
Therefore, the \hmol\ line may be induced by C-shock. 
According to the NIR spectroscopic survey of FUors \citep{connelley18}, the \hmol~2.1218~\um~emission line was not found in bonafide FUors but mostly found in FUor-like objects and peculiar objects. 
The majority of the FUor-like objects and peculiar objects in \citet{connelley18} are classified as Class I sources in \citet{connelley10}, which imply that they still have surrounding envelope material.

\section{Comparison with HBC 722} \label{sec:disccusion}
The analyses of disk features (Section 3.2) show that the optical spectrum traces warmer material at higher velocity (smaller radius) than the NIR spectrum traces. 
These results show evidence for Keplerian disk rotation which was also found in HBC~722 \citep{lee15}. 
Both of the FUors have pre-outburst data and have high-resolution spectroscopic monitoring data in the optical and NIR after their outburst.
Therefore, we compared the two FUors to characterize V960~Mon.
Table~\ref{tbl_comp} lists data useful for comparisons between V960~Mon and HBC~722.

The two FUors show disk features in the optical and NIR spectra, and the trend of velocity with wavelength is similar.
The optical and NIR spectra of HBC~722 trace the disk radius of about 39$\pm$7~\Rsun\ and 76$\pm$13~\Rsun\ at temperatures of about 5000~K and 3000~K, respectively, when the \Vmax\ \citep{lee15} is used.  
According to the SED modeling by \citet{gramajo14}, the inclination and mass of HBC~722 are 85 degrees (almost edge-on) and 1~\Msun, respectively. 
Therefore, the disk radii traced by the optical and NIR spectra and their corresponding temperatures in HBC~722, which were obtained using \Vmax\ and 1 \Msun\ \citep{lee15}, are adopted to compare with those of V960~Mon. 
The uncertainty of the estimated disk radius is calculated by error propagation adopting 10$\%$ error for mass and rotational velocity. 
The uncertainty of mass was adopted as the standard deviation of masses obtained by comparing the temperature and radius of HBC~722 presented in \citet{gramajo14} with three different evolutionary models \citep{siess00, bressan12, baraffe15}.
The standard deviation of the rotational velocities estimated in V960~Mon is about 10$\%$ of the mean rotational velocity for both optical and NIR. 
Since we applied the same technique to find the rotational velocity, we adopted the same uncertainty for the rotational velocity for HBC~722.
If we use the \Vmax\ of V960~Mon, the optical and NIR spectra trace a disk radius of about 88$\pm$32~\Rsun~and 109$\pm$40~\Rsun~with temperatures of about 5300-5000~K and 4600~K, respectively.
Therefore, V960~Mon is hotter than HBC~722 at the radii traced by the optical and NIR spectra if the maximum projected rotational velocities are adopted.

\begin{figure}
\epsscale{0.8}
\plotone{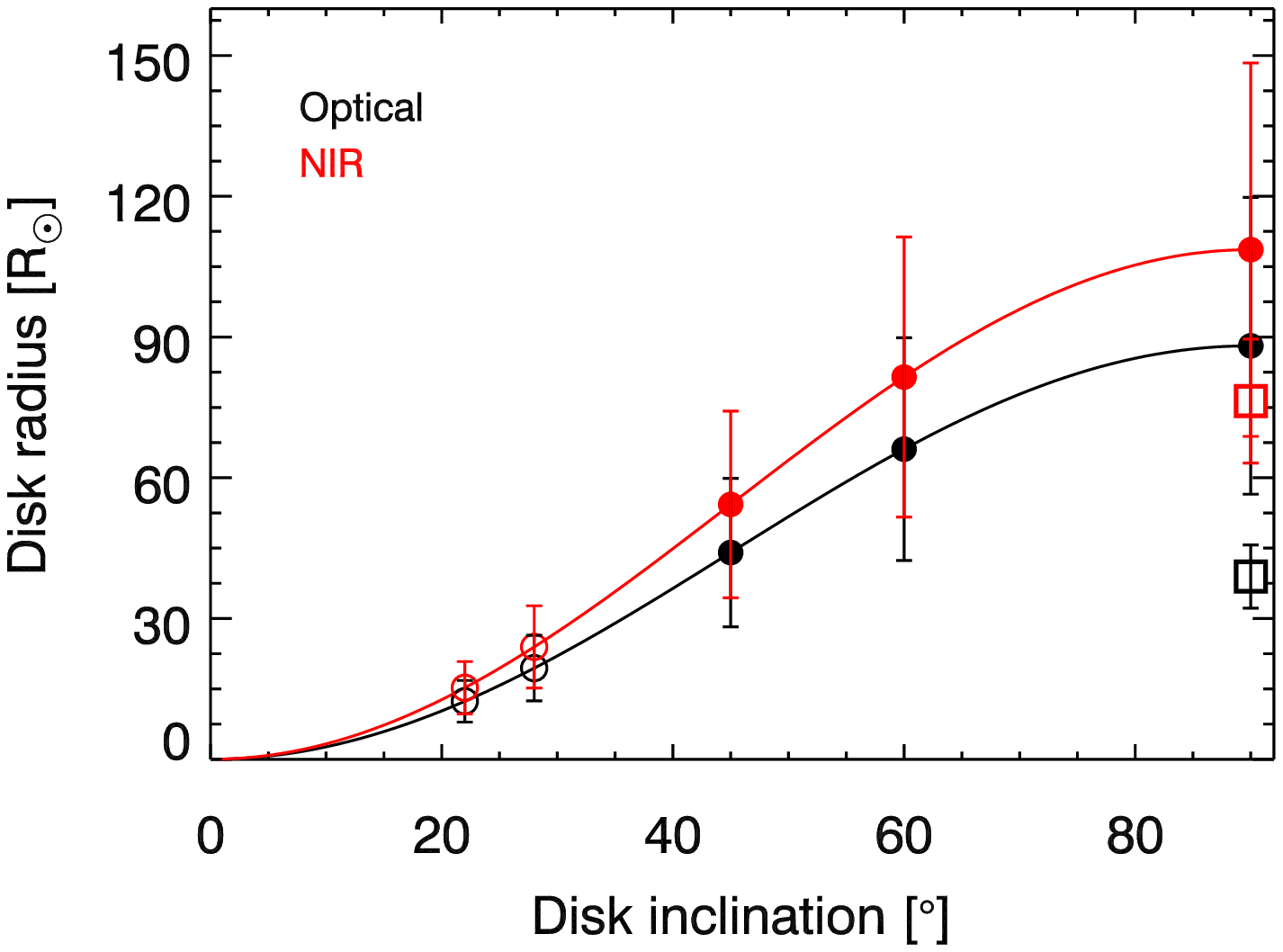} 
\caption{
Estimated disk radius as a function of disk inclination. 
Black and red lines indicate the calculated disk radius by the best-fit rotational velocity of optical (\Vmax\ = 40.3$\pm$3.8~\kms) and NIR (\Vmax\ = 36.3$\pm$3.9~\kms), respectively.
Circle and square symbols represent V960~Mon and HBC~722 \citep{lee15}, respectively.
The open circles indicate the disk radii corresponding to the inclination (22 and 28 degrees) suggested by \citet{caratti15}, and the filled circles denote the disk radii corresponding to the inclinations of 45, 60, and 90 degrees. 
The uncertainty of the disk radius is calculated by the error propagation.
\label{fig_radius}
}
\end{figure}

Fig.~\ref{fig_radius} shows the comparison of disk radius between V960~Mon (circle) and HBC~722 (square). 
The solid lines indicate the estimated disk radius of optical (black) and NIR (red) as a function of disk inclination by adopting the best-fit rotational velocity of V960~Mon.
When disk inclination is assumed as 45 (60) degrees, the optical and NIR spectra of V960~Mon trace a disk radius of about 44$\pm$16 (66$\pm$24)~\Rsun\ and 54$\pm$20 (81$\pm$30)~\Rsun, respectively. 
If disk inclination is about 22 (28) degrees \citep{caratti15}, the observed spectra trace a disk radius of about 12$\pm$4~(19$\pm$7)~\Rsun\ and 15$\pm$6~(24$\pm$8)~\Rsun, respectively.

The bolometric luminosity (\Lbol) at the outburst stage of V960~Mon is about 48~\Lsun~\citep{connelley18} while that of HBC 722 is about 8.7 to 17~\Lsun~\citep{kospal11, kospal16, connelley18} suggesting that the disk of V960~Mon can be hotter than that of HBC~722 at the outburst stage. 
Since the accretion luminosity (\Lacc) dominates the \Lbol\ in FUors \citep{hartmann96, konigl11}, \Lbol~is proportional to the mass accretion rate (\Macc): \Lbol~$\sim$~\Lacc~$\propto$~\textit{M}$\times$\Macc~\citep{hartmann96, hartmann09}. 
The higher \Lbol\ indicates higher \Macc, which implies that a massive accretion heating occurs and the disk midplane becomes hotter \citep[\Tdisk~$\propto$~\Macc$^{1/4}$,][]{zhu07, hartmann09}. 
In addition, the \Lbol~of V960~Mon \citep[4.8~\Lsun;][]{kospal15} at the pre-outburst stage is also higher than that of HBC 722 \citep[0.85~\Lsun;][]{kospal11}.

If the mass of the two FUors is similar to 1~\Msun~(see Table~\ref{tbl_comp}), then the relatively higher \Lbol~of V960~Mon means a relatively higher \Macc~than that of HBC~722.
Moreover, the upper limit of the disk mass of HBC~722 is about 0.01-0.02~\Msun~\citep{dunham12, kospal16} while the circumstellar mass of V960~Mon is about 0.01-0.06~\Msun~\citep{kospal16}. 
Since V960~Mon is known as Class II before its outburst \citep{kospal15}, we can assume that the circumstellar mass is dominated by the disk mass. 
Then, the disk masses of V960~Mon and HBC~722 are similar. 
Even if the protostellar masses and the disk masses are similar, the \Macc~of V960~Mon is higher than that of HBC~722.

Another difference between the two FUors is the existence of emission line.
The emission lines have been hardly detected in FUors, but we detected three emission lines of the \halp~6563~\AA{}, the [\SII]~6731~\AA{} and the \hmol~2.1218~\um~in V960 Mon. 
The [\SII]~6731~\AA{} and the \hmol~2.1218~\um\ emission lines are known as jet/outflow tracers.
The [\SII]~6731~\AA{} emission line is often detected in Class II \citep{hirth97, simon16}, while the \hmol~2.1218~\um~emission line in Class I \citep{davis03, bally07, bally16}.
However, from the SEDs of their pre-outburst stage, the two FUors are known as Class II \citep{kospal11, miller11, kospal15, kospal16}, and their spectral index ($\alpha$) is also about -0.4 \citep[HBC~722;][]{miller11} and -0.5 \citep[V960~Mon;][]{kospal15} which are the typical values of Class II.
In previous studies, HBC~722 is classified as an evolved Class II because outflow feature was not observed and its envelope mass is small \citep{green11, dunham12}.
Of the three emission lines detected in V960~Mon, only \halp\ P Cygni profile was detected in HBC~722.

The bolometric temperatures (\Tbol) is an evolutionary indicator \citep{myers93, chen95}, and the \Tbol~of V960~Mon in the pre-outburst stage was about 1190~K \citep{kospal15} while \Tbol~of HBC~722 in the pre-outburst phase is unknown. 
Therefore, we estimated the \Tbol~of HBC~722 at the pre-outburst stage by adopting the photometric data \citep{guieu09, rebull11, barentsen14}. 
The calculated \Tbol~of HBC~722 in the pre-outburst stage is about 1451 $\pm$ 11~K, higher than that of V960~Mon. 
The higher \Tbol~indicates a more evolved stage \citep{myers93}.
Therefore, the lower \Tbol~of V960~Mon than that of HBC~722 might indicate that V960~Mon is in a relatively earlier evolutionary stage than HBC~722 in the Class II stage.

\section{Conclusions} \label{sec:conclusion}
We have conducted monitoring observations of V960~Mon with high-resolution (R $\ge$ 30,000) optical and NIR spectrographs since 2014 December.
Several features of wind, disk, and outflow/jet were detected and enabled us to study the physical and kinematical properties of V960~Mon. 
From these data, we found the following results: \\
1. Wind features appear to vary with time.
The strength of the high-velocity wind feature was the strongest at the first observation, then continually weakened which can be explained by the decrease of the mass accretion rate. \\
2. The double-peaked absorption profiles of disk features are detected in both optical and NIR wavelengths. 
The spectral features of V960~Mon support the Keplerian disk rotation with a higher rotation velocity at the inner hotter disk, which is traced at a shorter wavelength. \\
3. The emission lines of the [\SII]~6731~\AA{} and the \hmol~2.1218~\um~are detected in our observations, which are rarely found in FUors. The \hmol~2.1218~\um~emission line in V960~Mon is detected in our observation for the first time. \\
4. The comparison with HBC~722, which is a more evolved Class II object according to its optical and NIR spectra, suggests that the disk of V960~Mon is probably hotter than that of HBC 722, and V960~Mon is in a relatively earlier Class II stage than HBC~722.

\clearpage

\acknowledgments
We acknowledge with thanks the variable star observations from the AAVSO International Database contributed by observers worldwide and used in this research.
This work was supported by the National Research Foundation of Korea (NRF) grant funded by the Korea government (MSIT) (grant numbers: NRF-2017R1A2B4007147, 2018R1A2B6003423, NRF-2019R1C1C1005224).
This work is also supported by the Korea Astronomy and Space Science Institute under the R\&D program supervised by the Ministry of Science, ICT and Future Planning.
This work used the Immersion Grating Infrared Spectrometer (IGRINS) that was developed under a collaboration between the University of Texas at Austin and the Korea Astronomy and Space Science Institute (KASI) with the financial support of the US National Science Foundation under grants AST-1229522 and AST-1702267, of the University of Texas at Austin, and of the Korean GMT Project of KASI. This paper includes data taken at The McDonald Observatory of The University of Texas at Austin. These results made use of the Discovery Channel Telescope at Lowell Observatory. Lowell is a private, non-profit institution dedicated to astrophysical research and public appreciation of astronomy and operates the DCT in partnership with Boston University, the University of Maryland, the University of Toledo, Northern Arizona University and Yale University.

\acknowledgments
\bibliographystyle{aasjournal}
\bibliography{ms_v960mon}


\begin{deluxetable}{cccccc}
\tabletypesize{\scriptsize} 
\tablecaption{Observation Logs of BOES and IGRINS \label{tbl_obs_tot}}
\tablewidth{0pt}
\tablehead{\colhead{Telescope} & \colhead{Instrument} & \colhead{Spectral Resolution} & \colhead{Observation Date} & \colhead{Exposure Time$^{a}$} &\colhead{Telluric Standard Star} \\
\colhead{} & \colhead{} & \colhead{} & \colhead{[UT]} & \colhead{[sec]} & \colhead{}}
\startdata
BOAO & BOES$^{b, \ast}$ & 30,000 & 2015 Feb 11 & 3600 & \\
$\cdots$ & $\cdots$ & $\cdots$ & 2015 Mar 08 & 3600 & \\
$\cdots$ & $\cdots$ & $\cdots$ & 2015 Mar 27 & 3600 & \\
$\cdots$ & $\cdots$ & $\cdots$ & 2015 Oct 10 & 3600 & \\
$\cdots$ & $\cdots$ & $\cdots$ & 2015 Dec 28 & 3600 & \\
$\cdots$ & $\cdots$ & $\cdots$ & 2016 Dec 16 & 3600 & \\ 
$\cdots$ & $\cdots$ & $\cdots$ & 2017 Mar 04 & 3600 & \\
$\cdots$ & $\cdots$ & $\cdots$ & 2017 Dec 28 & 3600 & \\
$\cdots$ & $\cdots$ & $\cdots$ & 2018 Mar 23 & 3600 & \\
$\cdots$ & $\cdots$ & $\cdots$ & 2018 Oct 08 & 3600 & \\
$\cdots$ & $\cdots$ & $\cdots$ & 2018 Dec 19 & 3600 & \\
 \cline{1-6}
HJST/McDonald & IGRINS$^{c}$ & 45,000 & 2014 Dec 25 & 400 (100 $\times$ ABBA) & HD 45380 \\ 
$\cdots$ & $\cdots$ & $\cdots$ & 2015 Jan 21 & 400 (100 $\times$ ABBA) & HD 53205 \\
$\cdots$ & $\cdots$ & $\cdots$ & 2015 May 02 & 1200 (300 $\times$ ABBA) & HD 53205 \\
$\cdots$ & $\cdots$ & $\cdots$ & 2015 Nov 23 & 960 (240 $\times$ ABBA) & HD 45380 \\
$\cdots$ & $\cdots$ & $\cdots$ & 2015 Nov 24 & 960 (240 $\times$ ABBA) & HIP 30594 \\
$\cdots$ & $\cdots$ & $\cdots$ & 2015 Dec 02 & 400 (100 $\times$ ABBA) & HD 53205 \\
$\cdots$ & $\cdots$ & $\cdots$ & 2016 Feb 28 & 480 (120 $\times$ ABBA) & HD 53205\\
DCT/Lowell & $\cdots$ & $\cdots$ & 2016 Nov 18 & 1200 (300 $\times$ ABBA) & HD 56525 \\
HJST/McDonald & $\cdots$ & $\cdots$ & 2017 Apr 17 & 1200 (300 $\times$ ABBA) & HR 2584 \\
DCT/Lowell & $\cdots$ & $\cdots$ & 2017 Nov 26 & 800 (200 $\times$ ABBA) & HR  1578 \\
\enddata
\tablenotetext{a}{Total integration time of each target (exposure time $\times$ the number of exposures = total integration time).}
\tablenotetext{b}{Wavelength coverage of BOES: 3900-9900 \AA{}}
\tablenotetext{c}{Wavelength coverage of IGRINS: H (1.49-1.80 \um) and K (1.96-2.46 \um) bands}
\tablenotetext{\ast}{HD 219477 (G2 II-III) and HD 18474 (G5 III) were observed as template spectra.}
\end{deluxetable}

\begin{deluxetable}{ccc}
\tabletypesize{\scriptsize}
\tablecaption{HWHD of Double-Peaked Lines\label{tbl_hwhd}}
\tablewidth{0pt}
\tablehead{\colhead{Wavelength} & \colhead{Element} & \colhead{HWHD} \\
\colhead{[\AA]} & \colhead{} & \colhead{[\kms]}}
\startdata
 5383 &   Fe I &  55.5 $\pm$  6.1 \\
 6411 &   Fe I &  45.4 $\pm$  7.8 \\
 6439 &   Ca I &  49.5 $\pm$  5.7 \\
 6449 &   Ca I &  51.4 $\pm$  4.8 \\
 6471 &   Ca I &  46.8 $\pm$  4.4 \\
15340 &   Fe I &  40.0 $\pm$  2.1 \\
15670 &   Fe I &  42.2 $\pm$  5.3 \\
16490 &   Fe I &  33.8 $\pm$  2.1 \\
22609 &   Ca I &  38.6 $\pm$  4.9 \\
\enddata
\end{deluxetable}

\begin{deluxetable}{ccccccc}
\tabletypesize{\scriptsize}
\tablecaption{Comparison \label{tbl_comp}}
\tablewidth{0pt}
\tablehead{\colhead{Instrument} & \colhead{Mass} & \colhead{Target} & \colhead{\Vmax} & \colhead{Spectral Type} & \colhead{Temperature} & \colhead{Radius$^\dagger$} \\
\colhead{} & \colhead{\Msun} & \colhead{} & \colhead{[\kms] } & \colhead{} & \colhead{[K]} & \colhead{[\Rsun]}}
\startdata
BOES 	    & 0.75 $\pm$ 0.25$^{1}$ & V960~Mon & 40.3$\pm$3.8 & G2 II-III / G5 III & 5308$^\ast$ / 5013$^{2}$  & 88$\pm$32 \\
	  	    & 0.8 $\sim$ 1.0$^{3, 4}$   		& HBC~722$^{\ast\ast}$ 			         & 70  & G5 II 	            & 5090$^{5}$  		 & 39$\pm$7 \\
IGRINS	    & 0.75 $\pm$ 0.25 		& V960~Mon & 36.3$\pm$3.9  & K1 III 	            & 4634$^{6}$   		 & 109$\pm$40 \\
		    & 0.8 $\sim$ 1.0   				& HBC~722 			        & 50  & K5 Iab 	            & 3055$^{7}$    		 & 76$\pm$13 \\
\enddata
\tablenotetext{\ast}{~Temperature is calculated by $T_\mathrm{eff}-(B-V)$ relation \citep{flower96, torres10}.} 
\tablenotetext{\ast\ast}{~The analytical data of HBC~722 are obtained from \citet{lee15}.} 
\tablenotetext{\dagger}{~Radius is obtained by using the maximum projected velocity (\Vmax).}
\tablerefs{
(1) \citet{kospal15}; (2) \citet{liu14}; (3) \citet{kospal16}; (4) \citet{gramajo14}; (5) \citet{kovtyukh07}; (6) \citet{wu11}; (7) \citet{bakos71}
}
\end{deluxetable}

\end{document}